\documentclass{article}
\usepackage[%
journal=IFB,    
lang=british,   
]{ems-journal}
\usepackage{comment}
\usepackage{mathtools}
\usepackage{amsmath}
\usepackage{makecell}
\usepackage{diagbox}
\usepackage{hyperref}

\numberwithin{equation}{section}

\begin{document}
\title{A topological approach to the Cahn-Hilliard equation and hyperuniform fields}

\emsauthor{1}{
	\givenname{Abel}
	\surname{H. G. Milor}
	\mrid{}
	\orcid{0009-0001-3890-0608}}{A.~Milor}
\emsauthor{2}{
	\givenname{Otto}
	\surname{Sumray}
	\mrid{1525220}
	\orcid{0000-0003-1735-0595}}{O.~Sumray}
\emsauthor{3}{
	\givenname{Heather A.}
	\surname{Harrington}
	\mrid{925307}
	\orcid{0000-0002-1705-7869}}{H.~Harrington}
\emsauthor{4}{
	\givenname{Axel}
	\surname{Voigt}
	\mrid{697268}
	\orcid{0000-0003-2564-3697}}{A.~Voigt}
\emsauthor{5}{
	\givenname{Marco}
	\surname{Salvalaglio}
	\mrid{1277852}
	\orcid{0000-0002-4217-0951}}{M.~Salvalaglio}

\Emsaffil{1}{
	\department{Institute of Scientific Computing}
	\organisation{Technische Universität Dresden}
	\rorid{042aqky30}
	\zip{01062}
	\city{Dresden}
	\country{Germany}
\affemail{abel$\_$henri$\_$guillaume.milor@tu-dresden.de}}
\Emsaffil{2}{
	\organisation{1}{Max Planck Institute of Molecular Cell Biology and Genetics}
	\rorid{042aqky30}
	\address{1}{Pfotenhauerstraße 108}
	\zip{1}{01307}
	\city{1}{Dresden}
	\country{1}{Germany}
	\affemail{sumray@mpi-cbg.de}
    \organisation{2}{Max Planck Institute for the Physics of Complex Systems}
    \address{2}{N\"othnitzer Straße 38}
	\zip{2}{01187}
	\city{2}{Dresden}
	\country{2}{Germany}
    }
\Emsaffil{3}{
	\organisation{1}{Max Planck Institute of Molecular Cell Biology and Genetics}
	\rorid{042aqky30}
	\address{Pfotenhauerstraße 108}
	\zip{1}{01307}
	\city{1}{Dresden}
	\country{1}{Germany}
	\affemail{harrington@mpi-cbg.de}
    \department{2}{Mathematical Institute}
    \organisation{2}{University of Oxford}
	\zip{2}{OX2 6GG}
	\city{2}{Oxford}
	\country{2}{UK}
    \department{3}{Cluster of Excellence Physics of Life (PoL)}
    \organisation{3}{Technische Universität Dresden}
	\zip{3}{01062}
	\city{3}{Dresden}
	\country{3}{Germany}
	}
\Emsaffil{4}{
	\department{1}{Institute of Scientific Computing}
	\organisation{1}{Technische Universität Dresden}
	\rorid{042aqky30}
	\zip{1}{01062}
	\city{1}{Dresden}
	\country{1}{Germany}
	\affemail{axel.voigt@tu-dresden.de}
    \department{2}{Cluster of Excellence Physics of Life (PoL)}
    \organisation{2}{Technische Universität Dresden}
	\zip{2}{01062}
	\city{2}{Dresden}
	\country{2}{Germany}
    \organisation{3}{Center for Systems Biology Dresden (CSBD)}
    \address{3}{Pfotenhauerstraße 108}
	\zip{3}{01307}
	\city{3}{Dresden}
	\country{3}{Germany}
	}

\Emsaffil{5}{
	\department{1}{Institute of Scientific Computing}
	\organisation{1}{Technische Universität Dresden}
	\rorid{042aqky30}
	\zip{1}{01062}
	\city{1}{Dresden}
	\country{1}{Germany}
	\affemail{marco.salvalaglio@tu-dresden.de}
	\department{2}{Dresden Center for Computational Materials Science}
	\organisation{2}{Technische Universität Dresden}
	\zip{2}{01062}
	\city{2}{Dresden}
	\country{2}{Germany}
	}

\classification[55N31]{35B36}

\keywords{Hyperuniformity, Topological Data Analysis, Cahn-Hilliard Equation, Gaussian Random Fields}

\begin{abstract} Hyperuniform structures are disordered, correlated systems in which density fluctuations are suppressed at large scales. Such a property generalizes the concept of order in patterns and is relevant across diverse physical systems. We present a numerical characterization of hyperuniform scalar fields that leverages persistent homology. Topological features across different lengths are represented in persistence diagrams, while similarities or differences between patterns are quantified through Wasserstein distances between these diagrams. We apply this framework to numerical solutions of the Cahn-Hilliard equation, a canonical model for generating hyperuniform scalar fields. We validate the approach against known features of the Cahn-Hilliard equation, including its scaling properties, convergence to the sharp interface limit, and self-similarity of the solutions. We then generalize the approach by studying Gaussian random fields exhibiting different degrees and classes of hyperuniformity, showing how the proposed approach can be exploited to reconstruct global properties from local topological information. Overall, we show how hyperuniform characteristics systematically correlate with distributions of topological features in disordered correlated fields. We expect this analysis to be applicable to a wide range of scalar fields, particularly those involving interfaces and free boundaries.
\end{abstract}

\maketitle

\section{Introduction}

The study of pattern formation is essential for understanding natural phenomena, revealing how order emerges from complex interactions \cite{Gierer1972,Langer1980,Cross1993pattern,falco2025nonlocal}. Ordered systems, such as periodic (crystalline) arrangements, are special situations. In most cases, emerging natural patterns retain some degree of disorder. \textit{Hyperuniformity} refers to a class of arrangements that bridges the gap between ordered and disordered systems \cite{torquato2003local,torquato2018hyperuniform}. In particular, disordered hyperuniform (HU) systems exhibit suppressed density fluctuations at large scales, similar to those found in ordered arrangements. However, they lack translational and rotational symmetry, characteristic of fully amorphous or liquid systems. Disordered HU systems thus combine the long-range order of crystals with the short-range isotropy of liquids. They have been found in biological \cite{JiaoPRE2014,mayer2015well,Zheng2024,Backofen24} and physical systems \cite{xie2013hyperuniformity,martelli2017largescale,salvalaglio2020hyperuniform}. Moreover, they have been exploited to tune material properties, e.g., in optics \cite{yu2021engineered, Vynck2023}. See \cite{torquato2018hyperuniform} for a comprehensive review. 

We focus here on patterns that emerge in two-phase systems, e.g., from spinodal decomposition, and study their local arrangements. The corresponding phase separation process can be qualitatively modeled by the Cahn-Hilliard equation \cite{Cahn1958}. As the equation does not impose periodicity or a fixed characteristic length scale, other than the width of the diffuse interface, the emerging patterns are disordered. However, they display and maintain hyperuniformity throughout the coarsening regime \cite{ma2017random}, thus representing a prominent example of disordered HU systems. The Cahn-Hilliard equation is therefore used to develop and test topological measures for characterizing HU scalar fields, which are potentially applicable to interface and free boundary problems in general.

The main motivation for inspecting local arrangements of HU fields lies in the limited description of these systems at small to intermediate length scales. Hyperuniformity is commonly described through global measures, such as the spectral density for scalar fields. Its behavior at infinity determines if the patterns are HU or not  \cite{torquato2018hyperuniform}. However, most systems of interest, resulting from experiments or numerical simulations, are finite; they may exhibit patterns similar to HU systems, although the definition of hyperuniformity does not hold rigorously. For this reason, various phenomenological measures have been proposed, which extrapolate the behavior towards infinity using the value at the largest available size \cite{ma2017random} or compare the quantities for different finite sizes \cite{Maher2024}. Rigorous significance tests for hyperuniformity have also been proposed \cite{klatt2022genuine}, addressing this issue. Other approaches overcome these limitations using topological measures. For HU point patterns, measures based on the statistics of local graph neighborhoods and \textit{persistent homology} have been introduced \cite{salvalaglio2024persistent}. For the former approach see \cite{lazar2012complete,lazar2015topological,Skinner2021,skinner2022topological}, for the latter  \cite{edelsbrunner2002topological, edelsbrunner2008persistent,otter2017roadmap,edelsbrunner2022computational}. Using such descriptions, it has been shown numerically that HU point patterns exhibit distinctive local arrangements, enabling the identification of traits of hyperuniformity in finite-sized systems and outlining general properties of the manifold of HU patterns \cite{salvalaglio2024persistent}. The description via persistent homology can also be used in inverse approaches \cite{Milor2025inferring}. The extension of this framework to HU fields, however, is missing. It is presented here to characterize two-phase systems with diffuse interfaces, exemplified by the Cahn-Hilliard equation, and disordered correlated fields in general through a convenient parametrization of Gaussian random fields.

This paper is organized as follows. We review the mathematical background in Sect.~\ref{sec:mathbackground} introducing the basic notion and needed concepts of hyperuniformity (Sect.~\ref{sec:HU}), the Cahn-Hilliard equation as well as the considered numerical approach (Sect.~\ref{sec:CH}), and persistent homology based on a filtration accounting for the signed distance to a level-set of the field variable (Sect.~\ref{sec:PH}). Results are reported in Sect.~\ref{sec:results}. We report first on the characterization of HU patterns resulting from numerical solutions of the Cahn-Hilliard equation in Sect.~\ref{sec:resCH}. The convergence of both the numerical solutions and topological measures with decreasing thickness of the diffuse interface is discussed. We also study the impact of the latter on the self-similarity of the solutions of the Cahn-Hilliard equation over time using topological data analysis. In Sect.~\ref{sec:resGRF}, we then generalize the study to patterns with different HU characters, realized via conveniently parametrized Gaussian random fields. We discuss how distributions of topological features in persistence diagrams correlate with the global HU character and demonstrate that their relations can be inverted. This latter aspect is addressed by comparing Wasserstein distances of persistent diagrams with reference ones. This comparison not only allows for reproducing the correct HU characteristics of the Cahn-Hilliard equation, but it is also expected to provide, in general, more accurate results than previous approaches when dealing with finite sizes. A proof of concept is also presented, demonstrating the inference of hyperuniformity in the considered Gaussian random fields using a neural network. Finally, our conclusions are summarized in Sect.~\ref{sec:conclusions}. 

\section{Mathematical background}
\label{sec:mathbackground}

\subsection{Hyperuniformity}
\label{sec:HU}

Hyperuniformity refers to the suppression of large-scale density fluctuations. A comprehensive review can be found in \cite{torquato2018hyperuniform}. Here, we recall only the main information concerning basic definitions and quantities required for our purpose. Consider a field $\varphi(\mathbf{x}) : \Omega \to \mathbb{R}$ with $\Omega \subset \mathbb{R}^d$. Its autocovariance reads
\begin{equation}
\psi(\mathbf{x}_1-\mathbf{x}_2) = 
\big\langle  \left( \varphi(\mathbf{x}_1)-\langle \varphi(\mathbf{x}_1\right) \rangle)  \left( \varphi(\mathbf{x}_2)-\langle \varphi(\mathbf{x}_2\right) \rangle) 
\big\rangle,
\end{equation}
with $\mathbf{x}_1, \mathbf{x}_2 \in \Omega$ and $\langle \ \cdot \ \rangle$ denoting the ensemble average. Its Fourier transform yields the spectral density
\begin{equation}\label{eq:spectr-density}
\widehat{\psi}(\varphi,\mathbf{k}) = |\widehat{\varphi}(\mathbf{k})|^2,
\end{equation}
with $\mathbf{k}$ being the wave vector.
The scalar field $\varphi(\mathbf{x})$ is said to be \emph{hyperuniform} if:
\begin{equation}\label{eq:def_field_hypruniformity}
\lim_{|\mathbf{k}| \to 0} \widehat{\psi}(\varphi,\mathbf{k}) = 0.
\end{equation}
Moreover, a field $\varphi(\mathbf{x})$ for which $\widehat{\psi}(\varphi,\mathbf{k}) \sim |\mathbf{k}|^\alpha$ for $|\mathbf{k}| \to 0$ is called class-I ($\alpha > 1$), class-II ($\alpha = 1$) and class-III ($0 < \alpha < 1$) HU. The notion of anti-hyperuniformity has been established for $\alpha < 0$.
Hereafter we denote $\widehat{\psi}(\varphi,\mathbf{k})\equiv \widehat{\psi}(\mathbf{k})$ implicitly referring to the scalar field under consideration and $k=|\mathbf{k}|$. We are primarily interested in the behavior of the
angular average of $\widehat{\psi}(\mathbf{k})$, which 
we simply denote as $\widehat{\psi}(k)$. 

Given $\varphi(\mathbf{x})$, a key aspect is thus determining $\alpha$ in $\widehat{\psi}(k) \sim k^\alpha$ for $k \to 0$. As the limit $k \to 0$ refers to infinite systems, the above definition of hyperuniformity is strictly speaking not applicable to finite systems. For these, however, the hyperuniformity metric 
\begin{equation}\label{eq:HH}
    H = \lim_{k \to 0} \frac{\widehat{\psi}(k)}{\widehat{\psi}(k_{\rm peak})} ,
\end{equation}
has been established, with $k_{\rm peak}$ corresponding to the peak of the spectra. For an ideal HU system, $H = 0$. However, in many systems, $H$ can be small but nonzero. In the literature, systems exhibiting $H<10^{-4}$ are often referred to as effectively HU, and $H<10^{-2}$ as nearly HU. In practice, the corresponding quantity accessible in finite systems is
\begin{equation}\label{eq:tildeHH}
    \widetilde{H} = \frac{\widehat{\psi}(k_{\rm min})}{\widehat{\psi}(k_{\rm peak})} ,
\end{equation}
with $k_{\rm min}$ the smallest accessible length of wavevectors given the system size. Note that, in general, $\widetilde{H}$ overestimates $H$.

\subsection{Cahn-Hilliard equation}
\label{sec:CH}

We consider the Cahn-Hilliard equation for the scalar (phase) field $\varphi : \Omega \times [0,T] \to \mathbb{R}$ with $\Omega \subset \mathbb{R}^d$:
\begin{equation}\label{eq:cahn-hilliard}
\frac{\partial \varphi}{\partial t} + \Delta \left( \epsilon \Delta \varphi - \frac{1}{\epsilon} W'(\varphi) \right) = 0 \quad \mbox{in } \Omega \times [0,T],
\end{equation}
with periodic boundary conditions and $\varphi = \varphi_0$ in $\Omega \times \{0\}$. Thereby, $W(\varphi)=18\varphi^2(1-\varphi)^2$ is a double-well potential with minima at $\varphi = 0$ and $\varphi = 1$ and $\epsilon > 0$ a small parameter determining the width of the diffuse interface. It is a well-established model for phase separation in two-phase systems, assuming isotropy and constant temperature. Here, the Cahn-Hilliard equation serves as a prototypical model for pattern formation, providing HU features \cite{ma2017random}. 

Several properties are known for the Cahn-Hilliard equation, see \cite{miranville2019cahn}. Within the coarsening regime, where two phases, one in which $\varphi \approx 0$ and one in which $\varphi \approx 1$, have formed and are separated by a diffuse interface, centered around $\varphi = 0.5$, the equation converges for $\epsilon \to 0$ to the Hele-Shaw problem \cite{P_1989,ABC_1994}.
Scale invariance of this sharp interface limit suggests that the characteristic length scale of the pattern grows proportionally to $t^{1/3}$. Besides this scaling law, solutions with random initial conditions are also believed to be statistically self-similar. These properties of the Cahn-Hilliard equation have been widely corroborated by numerical simulations. We will use them to validate our topological approach for characterizing the emerging patterns.

At equilibrium, the diffuse interface corresponds to 
\begin{equation}\label{eq:tanh}
\varphi = \frac{1}{2} \left(1 - \tanh\left( \frac{3d}{\epsilon} \right) \right),
\end{equation}
see e.g. \cite{li2009geometric}, with $d=\mathcal{D}(\varphi(\mathbf{x}),0.5)$ the signed distance from the 0.5 isoline of $\varphi$  defined by
\begin{equation}\label{eq:signed_distance}
\mathcal{D}(\varphi(\mathbf{x}),c) = 
\begin{cases}
-\mathrm{dist}(\mathbf{x}, L_c) & \text{if } \varphi(\mathbf{x}) > c \\
+\mathrm{dist}(\mathbf{x}, L_c) & \text{if } \varphi(\mathbf{x}) < c \\
0 & \text{if } \varphi(\mathbf{x}) = c 
\end{cases},
\end{equation}
with $L_c = \{ x \in \Omega : \varphi(\mathbf{x}) = c \} $, and $\mathrm{dist}(\mathbf{x}, L_c)$ the Euclidean distance of $\mathbf{x}$ from $L_c$. Within the coarsening regime, eq.~\eqref{eq:tanh} provides a good approximation of $\varphi(\mathbf{x})$ and is often used to simplify numerical schemes \cite{torabi2009new,salvalaglio2021doubly}.

Here, we use a pseudospectral Fourier method to solve eq.~\eqref{eq:cahn-hilliard}. For $\Omega = [0,L]^d$, a periodic domain, it is a convenient approach. Let $\widehat{\varphi}_\mathbf{q}(t)$ be the (discrete) Fourier transform of $\varphi(\mathbf{x},t)$. By applying the Fourier transform to eq~\eqref{eq:cahn-hilliard}, we obtain the following equation for $\widehat{\varphi}_{\mathbf{q}}(t)$,
\begin{equation}
\frac{\partial \widehat{\varphi}_{\mathbf{q}}}{\partial t} + |\mathbf{q}|^2\left(\epsilon |\mathbf{q}|^2 \widehat{\varphi}_{\mathbf{q}}  + \frac{18}{\epsilon}\left(2 \widehat{\varphi}_{\mathbf{q}} + 6(\widehat{\varphi^2})_{\mathbf{q}} - 4(\widehat{\varphi^3})_{\mathbf{q}} \right) \right) = 0, \qquad \forall \, \mathbf{q}
\label{eq:chF}
\end{equation}
with $\mathbf{q}$ the Fourier-space vectors. Approximating the time derivative using a first-order finite difference scheme $ (\widehat{\varphi}_k^{n+1} - \widehat{\varphi}_k^n)/\tau$, with timestep $\tau$, eq.~\eqref {eq:chF} can be solved for each time instance using a semi-implicit time discretization as follows
\begin{equation}
\widehat{\varphi}_{q}^{n+1}=\frac{\widehat{\varphi}_{\mathbf{q}}^n- \tau |\mathbf{q}|^2 \left( \frac{108}{\epsilon}\widehat{[(\varphi^{n})^2}]_{\mathbf{q}}-\frac{72}{\epsilon}[\widehat{(\varphi^{n})^3}]_{\mathbf{q}}\right)}{1+\tau \left( \epsilon |\mathbf{q}|^4 - \frac{36}{\epsilon} |\mathbf{q}|^2\right)} ,
\label{eq:chFscheme}
\end{equation}
where all linear terms are treated implicitly \cite{chen1998applications}. Several more advanced approaches are available \cite{eyre1998unconditionally,HE20095101} but are not needed for the purposes of this paper. $\varphi^{n+1}$ is determined as the inverse Fourier transform of $\widehat{\varphi_{\mathbf{q}}}^{n+1}$. The discretization of eq.~\eqref {eq:chFscheme} requires employing the discrete Fourier transform. For $d=2$ this is typically done considering a regular $N\times N$ grid of $\mathbf{q}$ points with spacing ${2\pi}/{L}$. This translates to a spatial discretization $h = L/N$.

\begin{figure}
    \centering
    \includegraphics[width=\textwidth]{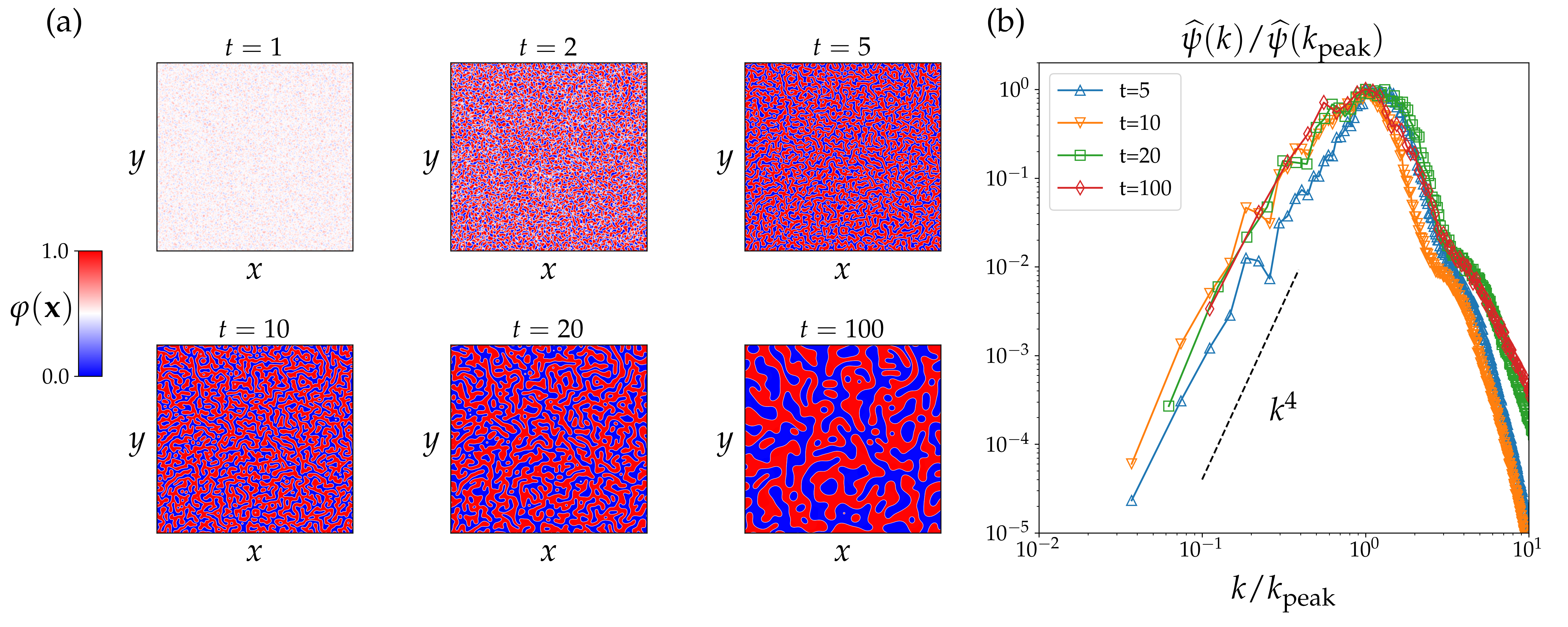}
    \caption{Evolving two-phase system governed by the Cahn--Hilliard equation \eqref{eq:cahn-hilliard} as a prototypical example of a HU system. (a) Numerical results obtained by integrating eq.~\eqref{eq:chFscheme} with $L=200$, $\epsilon=1$, $\tau=0.1$, and $N=2048$, corresponding to a spatial discretization $h\approx 0.1$. (b) Normalized spectral density $\widehat{\psi}(k)/\widehat{\psi}(k_{\rm peak})$, of representative patterns in panel (a) exhibiting a scaling $k^4$ for $k \to 0$ (within the coarsening phase).}
    \label{fig:figure1}
\end{figure}

Figure \ref{fig:figure1}(a) illustrates the evolution of an initially random field with a mean value of 0.5 dictated by eq~\eqref{eq:cahn-hilliard} solved as in eq.~\eqref{eq:chFscheme}. The angular averaged spectral density $\widehat{\psi}(k)$ for representative stages in the coarsening regime is reported in Figure~\ref{fig:figure1}(b). They indicate a scaling of $\widehat{\psi}(k) \sim k^4$ for $k \to 0$. Together with the values for $\widetilde H$, the lowest value of the normalized spectral density, this indicates effective hyperuniformity for $t=5$ and $t=10$ and nearly hyperuniformity for $t=20$ and $t=100$. The result is in agreement with previous computational results \cite{ma2017random,de2024hyperuniformity,padhan2025suppression}. However, the exponent $\alpha = 4$ also follows theoretically from the isotropy of the system at large spatial scales and directly links to the scaling of the characteristic length with $t^{1/3}$, see \cite{de2024hyperuniformity}.

\subsection{Persistent homology}\label{sec:PH}

Persistent homology (PH) is a central technique in topological data analysis (TDA) that quantifies the evolution of topological features in data across multiple scales (see, e.g., overviews in~\cite{edelsbrunner2022computational,otter2017roadmap,oudot2020inverse_persistence}).
Persistent homology takes as input a \emph{filtration}, an $\mathbb{R}$-indexed family of topological spaces $\{X_r\}_{r \in \mathbb{R}}$  such that for all $s \geq r $, we have $ X_r \subseteq X_s $. 
A standard construction of such a filtration is given by the sub-level sets of a scalar field $\varphi: \Omega \to \mathbb{R}$ on a domain $\Omega$,
where each $X_r$ is given by the subset 
\begin{equation}\label{eq:filtration_th}
X_r^{\rm th}:=\{ \mathbf{x} \in \Omega \, | \, \varphi(\mathbf{x})<r\}. 
\end{equation}
In practice, the domain $\Omega$ is discretised as a cubical complex with the filtration defined on the top-dimensional cells of the complex and extended to the lower-dimensional cells by taking the minimal value of their top-dimensional cofaces.

The nested structure of a filtration allows one to track the emergence and disappearance (i.e., \emph{birth} and \emph{death}) of $k$-dimensional homological features across the filtration, by computing the homology groups $ \mathcal{H}_k(X_r; \mathbb{F}) $, where $ \mathbb{F} $ is a field (commonly $ \mathbb{Z}_2$). The result is a \emph{persistence module}, 
which can be uniquely represented as a multiset of intervals $ \{[b_i, d_i]\} $ (often referred to as the \emph{barcode}), or the corresponding \emph{persistence diagram} $\mathcal{P}_k$ with points $ (b_i, d_i) \in \mathbb{R}^2 $, for each $\mathcal{H}_k$. Each bar $[b_i, d_i]$ corresponds to a $k$-dimensional homology class that first appears at birth index $b_i$,
and disappears at death index $d_i$.
Persistent homology thus provides a compact and multi-scale summary of the topological evolution of a space, capturing connected components ($\mathcal{H}_0$), loops ($\mathcal{H}_1$), voids ($\mathcal{H}_2$), and higher-dimensional features. The resulting barcodes or diagrams are stable under perturbations of the input and can be used as robust descriptors for pattern recognition, shape comparison, or analysis of complex data.

Persistent homology in the context of the Cahn-Hilliard equation using the sub-level filtration essentially operates only within the diffuse interface, as $\phi$ is constant within the two phases and $r \in [0,1]$. 
\begin{figure}
    \centering
    \includegraphics[width=\textwidth]{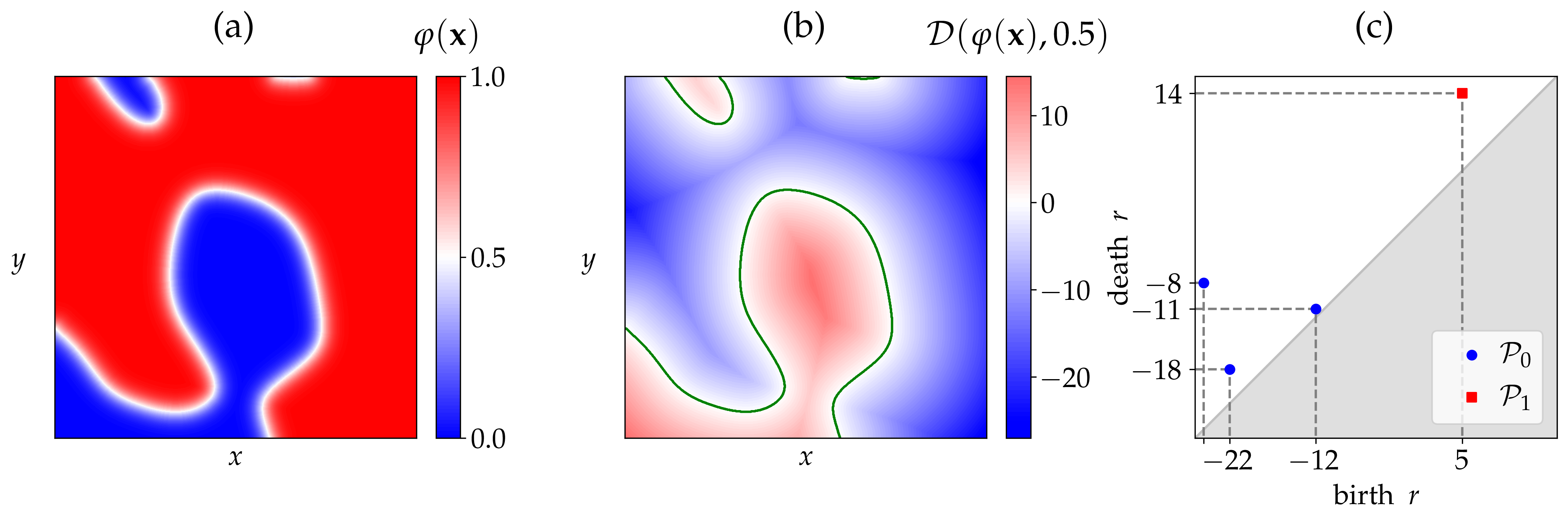}
    \includegraphics[width=\textwidth]{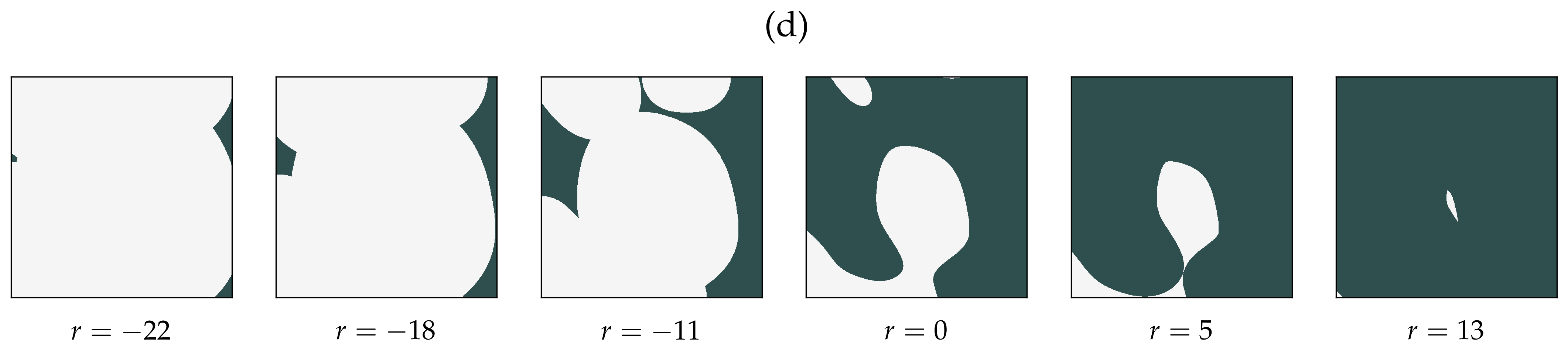}
    \caption{Illustration of the persistent homology of a two-phase system described by a smooth order parameter with diffuse interface among phases.
    (a) $\varphi(\mathbf{x})$ in a representative interface region from simulations in Figure \ref{fig:figure1}(a). Interface thickness $\epsilon$ is $1/10$ of the panel's linear dimension. (b) Signed distance $\mathcal{D}(\varphi(\mathbf{x}),0.5)$. (c) Persistence diagram obtained using the filtration \eqref{eq:filtration_iso}, further illustrated in panel (d). The latter displays the regions where $\mathcal{D}(\varphi(\mathbf{x}), 0.5) < r$ (dark grey) and $\mathcal{D}(\varphi(\mathbf{x}), 0.5) \geq r$ (light grey), for representative values of $r$ corresponding to, or close to, those defining the coordinates of the points in the persistence diagram shown in panel~(c).
    }
    \label{fig:persistent_homology}
\end{figure}
A richer filtration can be constructed by considering thresholds of the \textit{signed distance function} defined in eq.~\eqref{eq:signed_distance}, reading
\begin{equation}\label{eq:filtration_iso}
X_{r,c}^{\rm iso} = \{ \mathbf{x} \in \Omega : \mathcal{D}(\varphi(\mathbf{x}),c) \leq r \}.
\end{equation}
As $r$ increases from negative to positive, the sub-level sets sweep through the domain starting from regions below the isoline (e.g, valleys), passing through the isoline itself, and extending to regions above it (e.g., hills). For $c = 0.5$, $r < 0$ corresponds to regions of the phase $\phi = 0$, $r = 0$ includes the interface $\phi = 0.5$, and $r > 0$ adds regions of the phase $\phi = 1$, as can be readily seen from eq.~\eqref{eq:tanh}, too. The required values of $r$ thus no longer directly depend on the properties of $\varphi$ but depend on the considered isoline $L_c$ in the definition of $\mathcal{D}$ and the domain size. The filtration operates on the whole domain with $r \in [-\sqrt{2} L, \sqrt{2} L]$.
Crucially, the signed distance filtration also contains geometric data of each phase, such as `bottlenecks' in the isoline. 
Persistent homology of the signed distance function has been used to analyse porous materials such as bone vasculature~\cite{song2025generalized} and bone microstrutures~\cite{pritchard2023persistent}.
Applied to two-phase systems, this approach can thus characterize differences and similarities in the topological features associated with smaller variations in the interface. 

Figure~\ref{fig:persistent_homology}(a) illustrates $\varphi(\mathbf{x})$ in a small portion of the numerical solution of the Cahn-Hilliard equation from Figure~\ref{fig:figure1}. 
$\mathcal{D}(\varphi(\mathbf{x}),0.5)$ is shown in
Figure~\ref{fig:persistent_homology}(b). The filtration \eqref{eq:filtration_iso} is illustrated in Figure~\ref{fig:persistent_homology}(d) and delivers the persistent diagrams presented in Figure~\ref{fig:persistent_homology}(c).

To quantify differences between two persistence diagrams $\mathcal{P}_k$ and $\mathcal{P}_k^\prime$ we consider the $(p,q)$-Wasserstein distance \cite{skraba2020wasserstein, aktas2019persistence}
\begin{equation}\label{eq:wasserstein-distance}
W_{p,q}^k(\mathcal{P}_k,\mathcal{P}_k^\prime)=\inf_\gamma \bigg( \sum_{\mathbf{x} \in \mathcal{P}_k} (||\mathbf{x}-\gamma(\mathbf{x})||_q)^p\bigg)^{1/p} ,
\end{equation}
with $\gamma(\mathbf{x})$ ranging over all bijections from $\mathcal{P}_k$ to $\mathcal{P}_k^\prime$ and $|| \cdot ||_q$ the $L^q$ norm. When $\mathcal{P}_k$ and $\mathcal{P}_k^\prime$ contain different numbers of points, the matchings  $\gamma$ include distances to the persistence-diagram diagonal for the unmatched (supernumerary) elements. In the following, we will just consider the sum $W_{p,q}=W_{p,q}^0+W_{p,q}^1$ (known as the total Wasserstein distance) to characterize differences accounting for different homology groups in a scalar quantity. We note, however, that in all the cases reported below, $W_{p,q}^0$ and $W_{p,q}^1$ deliver similar values.
Persistent homology is stable with respect to this metric -- 
the total $(p,q)$-Wasserstein distance
between persistence diagrams constructed from scalar fields $\varphi, \varphi'$
has upper bound proportional to $\|\varphi - \varphi'\|_p$ (see~\cite{skraba2020wasserstein}).

\section{Results}
\label{sec:results}

\subsection{Cahn-Hilliard equation}
\label{sec:resCH}

\begin{figure}
    \centering    \includegraphics[width=1.0\textwidth]{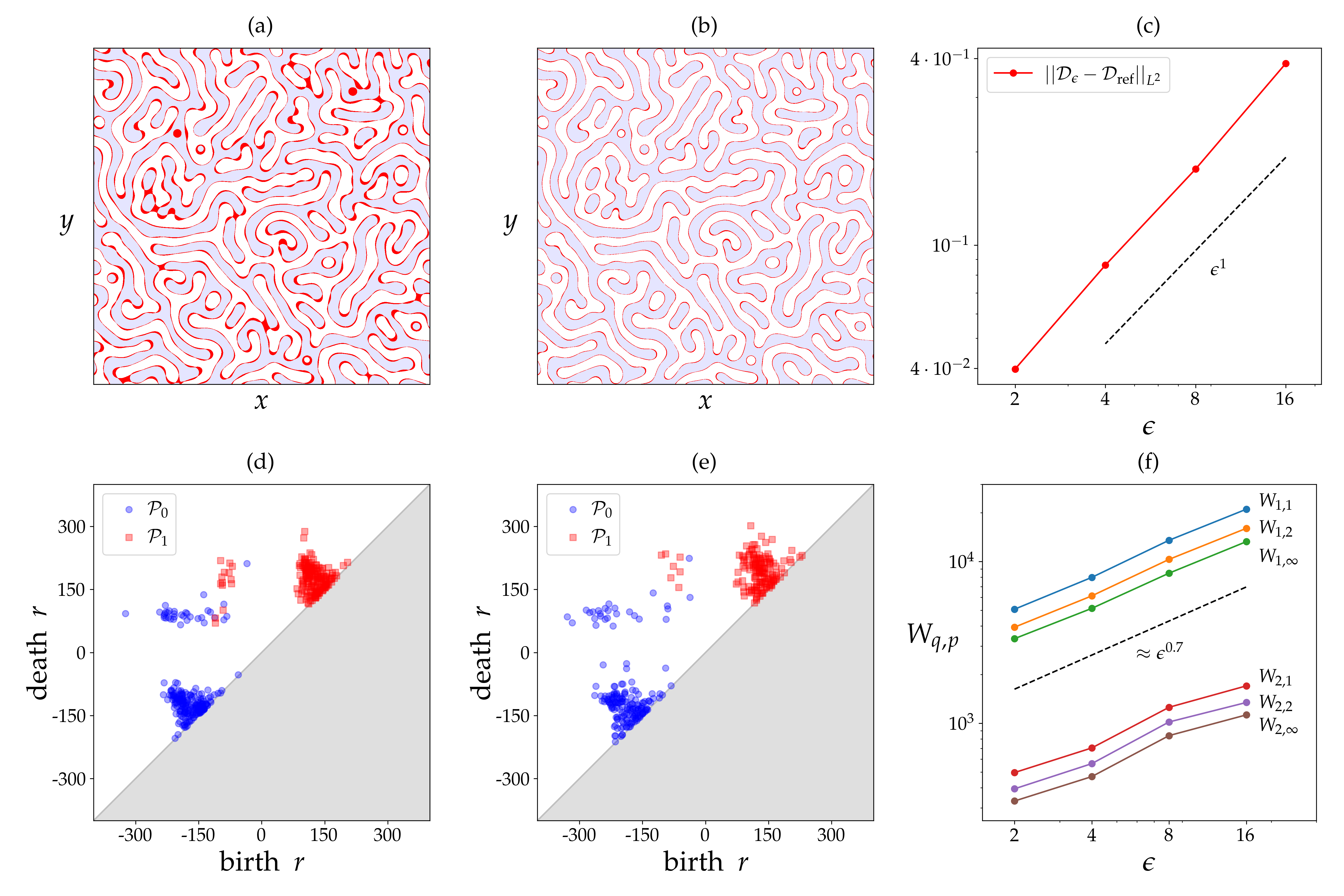}     \caption{Influence of $\epsilon$ on spinodal patterns obtained by solving the Cahn--Hilliard equation using $L=400$, $h\approx 0.1\epsilon$, $\tau=0.1$. (a),(b) Morphological differences at $t=50$ for varying $\epsilon$, highlighting deviations from a reference simulation ($\epsilon=0.5$, $\tau=0.01$, $h \approx 0.05$) via binarized domains ($\varphi>0.5$, light blue) and differences (red) for $\epsilon=8$ and $\epsilon=2$, respectively. (c) $L^2$ norm of the difference in signed distance from the $\varphi=0.5$ isoline indicates first-order convergence as $\epsilon \to 0$. (d),(e) Persistence diagrams computed from the filtration \eqref{eq:filtration_iso}, based on the signed distance \eqref{eq:signed_distance} to the $\varphi = 0.5$ isoline, for the patterns at $t = 50$ with $\epsilon = 8$ and $\epsilon = 2$, respectively. (f) Wasserstein distance $W_{q,p}$ between the persistence diagrams shown in panels (d) and (e), computed for varying $\epsilon$ with respect to the reference. An average sublinear convergence with scaling $\approx\epsilon^{0.7}$ is obtained.}
    \label{fig:figure3}
\end{figure}

Patterns corresponding to solutions of the Cahn--Hilliard equation \eqref{eq:cahn-hilliard} exhibit distinctive spatial organizations.  
Notably, both the morphology and connectivity of these patterns evolve over time and are sensitive to model parameters. PH offers a robust framework for quantifying and characterizing such geometric features across multiple spatial scales.

\subsubsection{Convergences in $\epsilon$}

We investigate the influence of the interface width ($\epsilon$) on the evolution of these patterns. To ensure comparable initial conditions, we enforce the same isoline $\varphi = 0.5$ (rather than starting from a random field with small amplitude as in Figure~\ref{fig:figure1}). To this end, we start from an initial pattern $\bar \varphi$ generated as in Figure~\ref{fig:figure1} with $\epsilon = 8$ and compute the signed distance $\mathcal{D}(\bar{\varphi},0.5)$ via eq.~\eqref{eq:signed_distance}. We then consider initial conditions $\varphi_{\epsilon}$ varying $\epsilon$ via the expression \eqref{eq:tanh} with $d=\mathcal{D}(\bar{\varphi},0.5)$. Note that the isoline $\varphi_{\epsilon}(\mathbf{x}, 0)=0.5$ does not depend on $\epsilon$ although the subsequent dynamics are affected by its value.

We use the pattern at $t=50$ obtained with $\epsilon = 0.5$ as reference ($\varphi_{\rm ref}$). Figures~\ref{fig:figure3}(a) and \ref{fig:figure3}(b) compare morphologies at $t=50$ for two different $\epsilon$ via binarization: the light blue regions indicate $\varphi_\epsilon > 0.5$, while the superposed red regions correspond to $|\chi(\varphi_\epsilon > 0.5) - \chi(\varphi_{\text{ref}} > 0.5)|$ with $\chi$ the characteristic function, highlighting differences between $\varphi_{\rm \epsilon}$ and the reference pattern. This already qualitatively illustrates a reduction in morphological differences as $\epsilon$ decreases. For a quantitative measure, Figure~\ref{fig:figure3}(c) reports the $L^2$ norm of the difference in signed distance fields from the $\varphi = 0.5$ isoline between $\varphi_\epsilon$ and the reference. A first-order convergence rate is obtained, which is comparable with results known for the Cahn--Hilliard equation approaching the Hele–Shaw problem as $\epsilon \to 0$ \cite{ABC_1994, feng2005numerical}.

\begin{figure}
    \centering
    \includegraphics[width=0.98\textwidth]{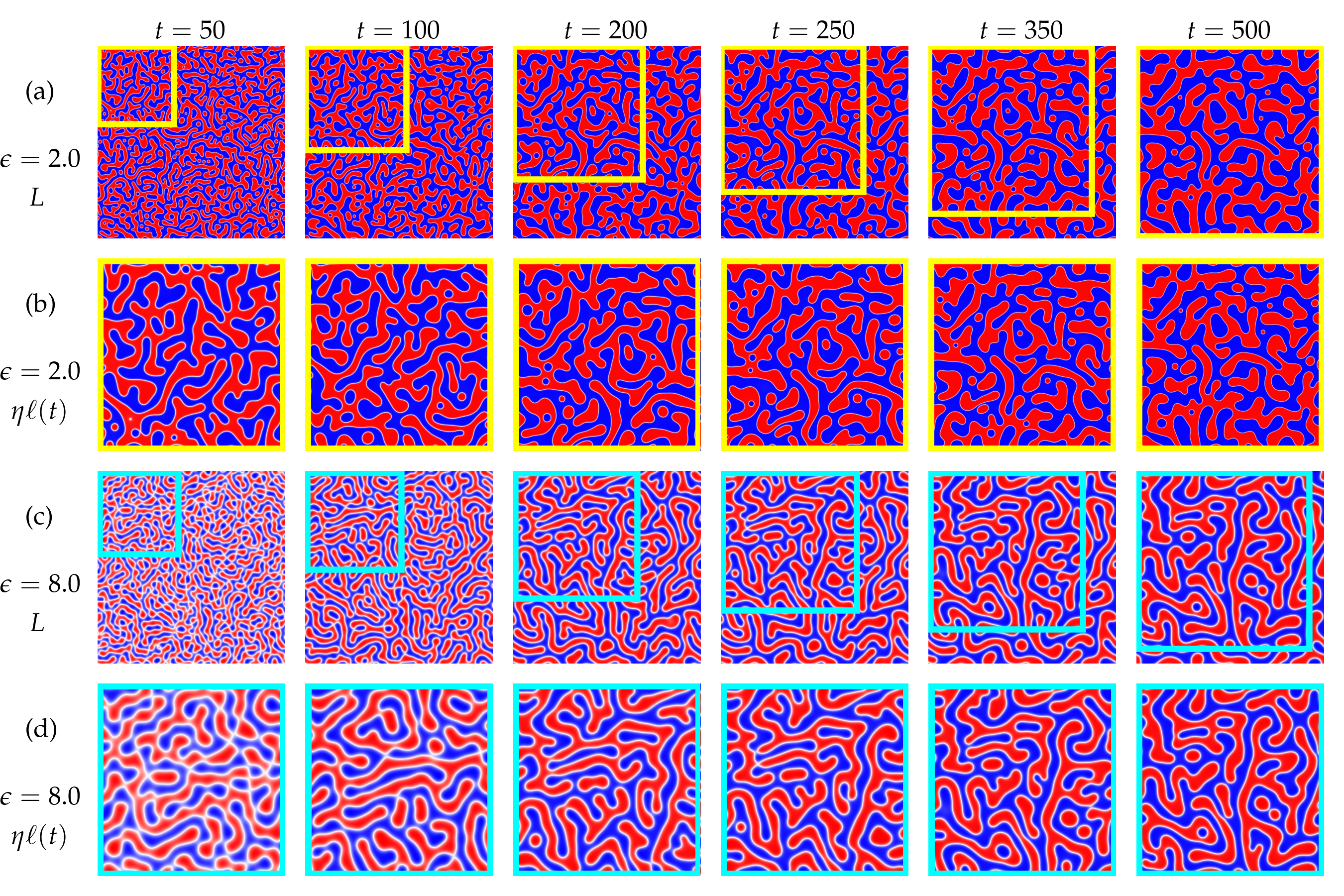}
    \includegraphics[width=0.98\textwidth]{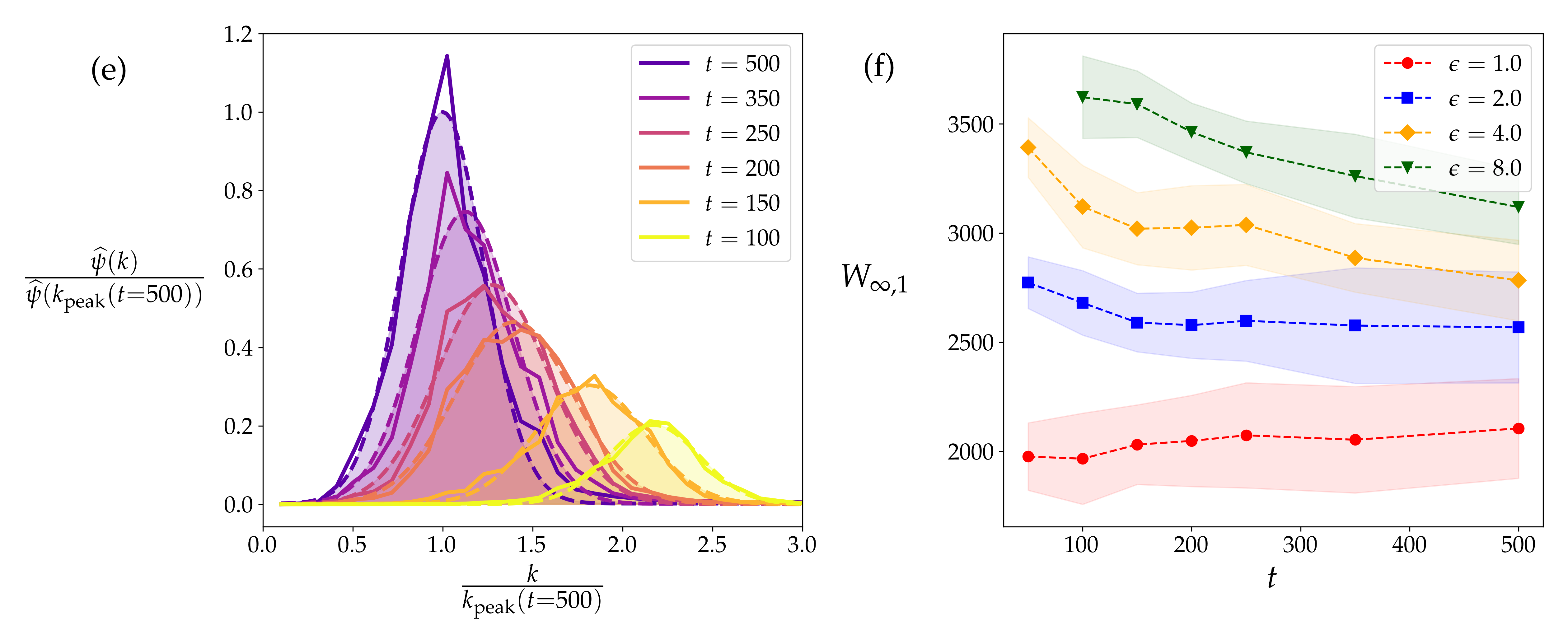}
    \caption{Topological assessment of self-similarity in numerical solutions of the Cahn-Hilliard equation. (a) Representative simulation stages (times $t$ indicated) for $\epsilon=2.0$, $L=400$; numerical parameters as in Figure~\ref{fig:figure1}. (b) Portions of size $\eta \ell(t) \times \eta \ell(t)$ with $\eta \sim 9$ extracted from panel (a) (yellow squares). (c) Same as panel (a) for $\epsilon=8$. (d) Portions of size $\eta \ell(t) \times \eta \ell(t)$ with $\eta \sim 9$ extracted from panel (c) (cyan squares). (e) Spectral density of the patterns in panel (d) (solid dots) with Gaussian fits (dashed lines, shaded areas). (f) Wasserstein distance $W_{\infty,1}$ \eqref{eq:wasserstein-distance} between persistence diagrams of portions of size $\eta \ell(t) \times \eta \ell(t)$, varying $\epsilon$ and time $t$. Symbols and shaded areas illustrate the average and standard deviation over 20 independent simulations ($t=50$ for $\epsilon=8.0$ is omitted, as it lies at the transition between the transient and coarsening regimes, see also panels (c),(d)).
    }
    \label{fig:figureSelf}
\end{figure}

Analysing the numerical simulation of the Cahn--Hilliard equation through PH allows for characterizing in detail how topological features are distributed, beyond global descriptors such as the spectral density $\psi(k)$ as in Figure~\ref{fig:figure1}(a) or the measure reported in Figure~\ref{fig:figure3}(c). Figures~\ref{fig:figure3}(d),(e) show persistence diagrams computed from the filtration \eqref{eq:filtration_iso} for the patterns at $t = 50$ with $\epsilon = 8$ and $\epsilon = 2$, respectively. Distinct clusters of points can be identified therein. Notably, the distributions in $\mathcal{P}_0$ and $\mathcal{P}_1$ are similar with respect to their birth and death $|r|$ values. This evidence directly reflects the underlying symmetry in the distribution of the two phases $\varphi_{\rm \epsilon}>0.5$ and $\varphi_{\rm \epsilon}<0.5$, which was set by the initial condition.

When varying $\epsilon$, differences are observed in the persistence diagrams; see Figure~\ref{fig:figure3}(d),(e). Wasserstein distances \eqref{eq:wasserstein-distance} for patterns obtained with varying $\epsilon$ with respect to the reference ($\epsilon=0.5$) are reported in Figure~\ref{fig:figure3}(f). They quantify differences qualitatively observed in Figure~\ref{fig:figure3}(a),(b) and globally characterized in Figure~\ref{fig:figure3}(c). Importantly, such distance measures also exhibit convergence as $\epsilon$ decreases, which at least qualitatively validates the approach. The convergence is sublinear, with a fitted average exponent of approximately 0.7. A convergence analysis for this measure is currently out of reach and remains an open problem to be investigated within TDA.

\subsubsection{Self-similarity}
The second feature of the Cahn-Hilliard equation, which can be used to validate our topological approach, is the self-similarity of the emerging patterns in the coarsening regime. We consider this feature for numerical simulations of the Cahn--Hilliard equation \eqref{eq:cahn-hilliard} starting from a random pattern with $\epsilon \in [1,8]$, $L=400$ over $t\in [0,500]$. At different stages, the resulting two-phase patterns exhibit structures with characteristic length $\ell(t)=2\pi/k_{\rm peak}(t)$ increasing over time, with $k_{\rm peak}(t)$ the position of the peak in the spectral density. To assess self-similarity over time, we can compare portions of the simulated patterns in a region $\eta \ell(t) \times \eta \ell(t)$ with $\eta \gg 1$. In the reported results, we consider $\eta \sim 9$ (at $t=500$, $9\ell \approx L$). Examples of the numerical simulations and the extracted portions to assess self-similarity are provided in Figure~\ref{fig:figureSelf}(a)--(d) for $\epsilon=2$ and $\epsilon=8$. The spectral densities of the reported snapshots for the simulations with $\epsilon=8$ are also illustrated in Figure~\ref{fig:figureSelf}(e). To improve its estimation, $k_{\rm peak}(t)$ (and thus $\ell(t)$) is obtained as the peak position of a Gaussian fit to the pattern’s spectral density, also illustrated therein.

Figure~\ref{fig:figureSelf}(f) illustrates the $W_{\infty,1}$ distance \eqref{eq:wasserstein-distance} between the persistence diagrams of a reference pattern ($\epsilon=0.5$, $t=500$) and those from simulations with different $\epsilon$ values over times, in all cases considering portions scaled by $\ell(t)$ as described above. The average (symbols) and standard deviation (shaded area) of the results of $W_{\infty,1}$ over 20 simulations are reported. We found that for small $\epsilon$ values ($\epsilon=1$), the distance varies within a very limited range, indicating effective self-similarity. By increasing $\epsilon$, we obtain that at early stages the distance slightly decreases over time while reaching an almost constant value ($\epsilon=2$). 
Increasing the interface thickness further leads to a more significant variation of the distance, retaining a generally decreasing trend. This behavior can be ascribed to the effect of a relatively large $\epsilon$ compared to $\ell(t)$ for early times. 

It is worth noting that, at later stages, $W_{\infty,1}$ in Figure~\ref{fig:figureSelf}(f) exhibits a decreasing trend with decreasing $\epsilon$., consistent with the result shown in Figure~\ref{fig:figure3}(f). Importantly, this result is nontrivial, as it is obtained by a quantitative evaluation of topological features across different patterns generated from random initial conditions, rather than from the same initial condition resolved with varying diffuse interface thicknesses as in Figure~\ref{fig:figure3}(f). Our topological approach thus not only qualitatively reproduces self-similarity for solutions of the Cahn-Hilliard equation, it also points to a convergence of this property for $\epsilon \to 0$, for which self-similarity is actually known. Again, any convergence analysis for the considered measure is beyond the scope of this paper.

\subsection{Hyperuniform Gaussian random fields}
\label{sec:resGRF}

In his seminal work on phase separation, J. Cahn \cite{Cahn1961} showed that the solution of the Cahn-Hilliard equation can be approximated by Gaussian random fields (GRFs).  
A Gaussian random field can be constructed as the superposition of a large number of plane waves with randomly distributed amplitudes and phases. This construction can be tailored to ensure that the field retains targeted statistical properties, such as isotropy and a specified correlation function. GRFs have been used to describe a vast amount of different structures in nature, geography, and cosmology, as well as in machine learning \cite{Hristopulos2020}. In our context of two-phase systems, GRFs with selected distribution of wavevectors have been exploited to design spinodal-like (spinodoid) patterns with tailored mechanical properties, e.g. \cite{Kumar2020,rosenkranz2025dataefficientinversedesignspinodoid}. We follow this approach to parameterize long-range correlated fields. Addressing the HU character of these fields will allow us to obtain the whole spectrum of HU characters. Considering the Wasserstein distances \eqref{eq:wasserstein-distance} between their persistence diagrams allows us to distinguish them, and when compared with new samples, to classify the patterns based on their HU characteristics.  

\subsubsection{Parametrization of HU scalar fields}

Consider fields
\begin{equation}\label{eq:GRF}
    \varphi(\mathbf{x})=\nu \sum_{n=1}^{N}A(k_n)\cos(\mathbf{x}\cdot\mathbf{k}_n+\phi_n),
\end{equation}
with $\mathbf{k}_n$ (wave)vectors with ideally random orientation and uniformly distributed lengths, $\phi_n$ random scalars in range $[0, 2\pi]$, 
$\nu$ a scaling factor ensuring $\max{|\varphi(\mathbf{x})|}=1$. We consider a square domain of size $L\times L$ and enforce periodicity of the fields. This is obtained by replacing $\mathbf{k}_n$ vectors with the nearest one with components $2\pi n/ L$ with $n$ an integer. As such fields are completely described by their spectral density $\widehat \psi(k_n)$, we can tune the amplitudes $A(k_n)$ to enforce a prescribed 
$\widehat \psi(k_n)$, given the relation $A^2(k_n)=\widehat \psi(k_n)$ following from \eqref{eq:spectr-density}. The spectral density is then defined including three tunable parameters $\alpha$, $H$, and $K$ as follows: 
\begin{equation}\label{eq:GRFparameters}
    \widehat \psi(k)=
    \begin{cases}
\left(\frac{k}{K}\right)^{\alpha}(1-H)+H & k \leq K, \\[6pt]
1 & K < k \leq 0.8 f_{\rm max}, \\[6pt]
e^{-10(0.8f_{\rm max}-k)^2} & 0.8 f_{\rm max} < k,
\end{cases}
\end{equation}
with $f_{\rm max}=\pi / (100L)$. $\alpha$ controls the exponent for $k \to 0$, and thus determines class-I HU, class-II HU, or class-III HU (or even non-HU or anti-HU) characteristics. $H$ directly corresponds to the quantity defined in \eqref{eq:HH}. $K$ determines the frequency (wavelength) below (above) which correlations are imposed.

\begin{figure}
    \centering
\includegraphics[width=1.0\textwidth]{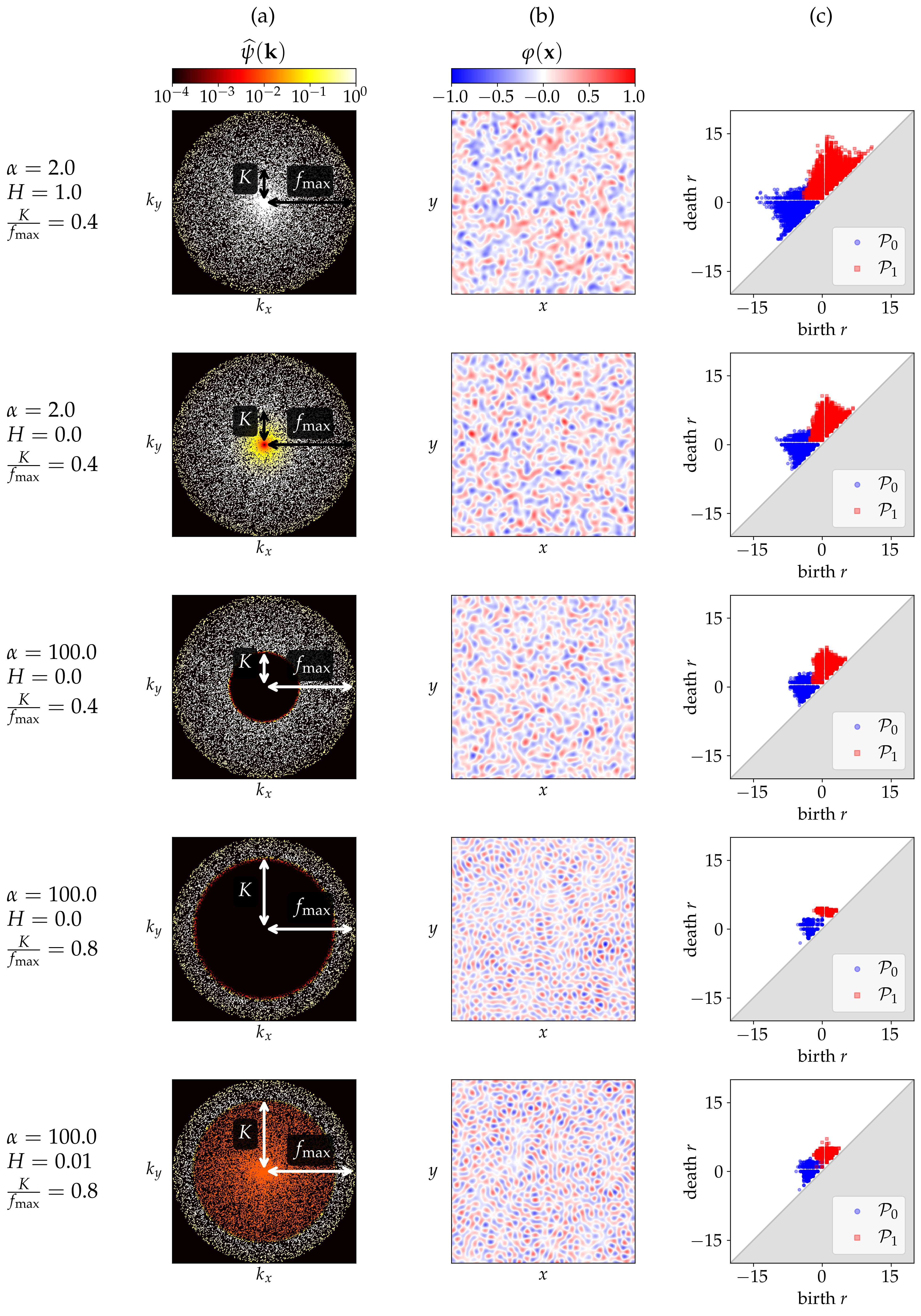}
    \caption{Generated GRFs with different HU or non-HU characters set via $\alpha$, $H$ and $K$. Five examples are shown, which are illustrated by: (a) The spectral density $\widehat{\psi}(\mathbf{k})$. The discrete character of the values for which $\widehat{\psi}(\mathbf{k})>0$ corresponds to the discrete sampling of the $\mathbf{k}$ vectors in eq.~\eqref{eq:GRF}. (b) Corresponding fields $\phi(\mathbf{x})$ from eq.~\eqref{eq:GRF}. Only a portion of the actual field is presented (1/64) for ease of visualisation. (c) Persistence diagrams computed via filtration \eqref{eq:filtration_iso} for each field in (b).}
\label{fig:figure4}
\end{figure}

With this parametrization, we generated different scalar fields using combinations of the values: 
\begin{equation*}
    \begin{split}
        \alpha&\in \{ 0;0.5;1;1.5;2;3;5;10;20;100\},\\
        H&\in\{0;10^{-4};10^{-3};10^{-2};10^{-1};0.35;1\},\\
        K/f_{\rm max}&\in\{0.32;0.40;0.48;0.56;0.64;0.72;0.80\}.
    \end{split}
\end{equation*}
For each combination, three different realizations are computed, resulting in a total of 1470 patterns. Examples of different prescribed sets of $\mathbf{k}_n$ vectors and amplitudes are illustrated in Figures~\ref{fig:figure4}(a) via the spectral density. Portions (1/64) of the corresponding generated fields $\varphi({\mathbf{x}})$ are shown in Figure~\ref{fig:figure4}(b). Persistence diagrams $\mathcal{P}_k$ computed leveraging the filtration $X^{\rm iso}_{r,c}$ with $c=0$ are shown in Figure~\ref{fig:figure4}(c). Note that for patterns obtained by the Cahn-Hilliard equations, we considered $c=0.5$. This difference arises from the definition of the GRF \eqref{eq:GRF}, which is bounded to $[-1,1]$ rather than $[0,1]$. Equivalent values can, however, be obtained through a trivial rescaling of \eqref{eq:GRF}, which, crucially, does not alter the signed distance from the chosen isoline corresponding to the rescaled value. Persistence diagrams exhibit similar properties to those obtained for the actual solution of the Cahn-Hilliard equation; see Figure~\ref{fig:figure3}. For instance, similar symmetry is retained concerning both $\mathcal{H}_0$ and $\mathcal{H}_1$ homological features. Broader point clouds are, however, obtained, pointing to a richer landscape of local arrangements.

It is worth noting that, for the GRFs defined in eq.~\eqref{eq:GRF}, the filtration \eqref{eq:filtration_th} also provides meaningful descriptions. As shown by the patterns in Figure~\ref{fig:figure4}, they exhibit more significant variation in the values of the corresponding scalar fields than within the diffuse interface of the Cahn-Hilliard equation considered in Sect.~\ref{sec:CH}. Nevertheless, in our tests (comparisons not shown), \eqref{eq:filtration_iso} provided better performance in all the characterizations and results reported below.

\subsubsection{Distances between patterns via persistent homology}
\label{sec:distance}

\begin{figure}
    \centering
    \includegraphics[width=\textwidth]{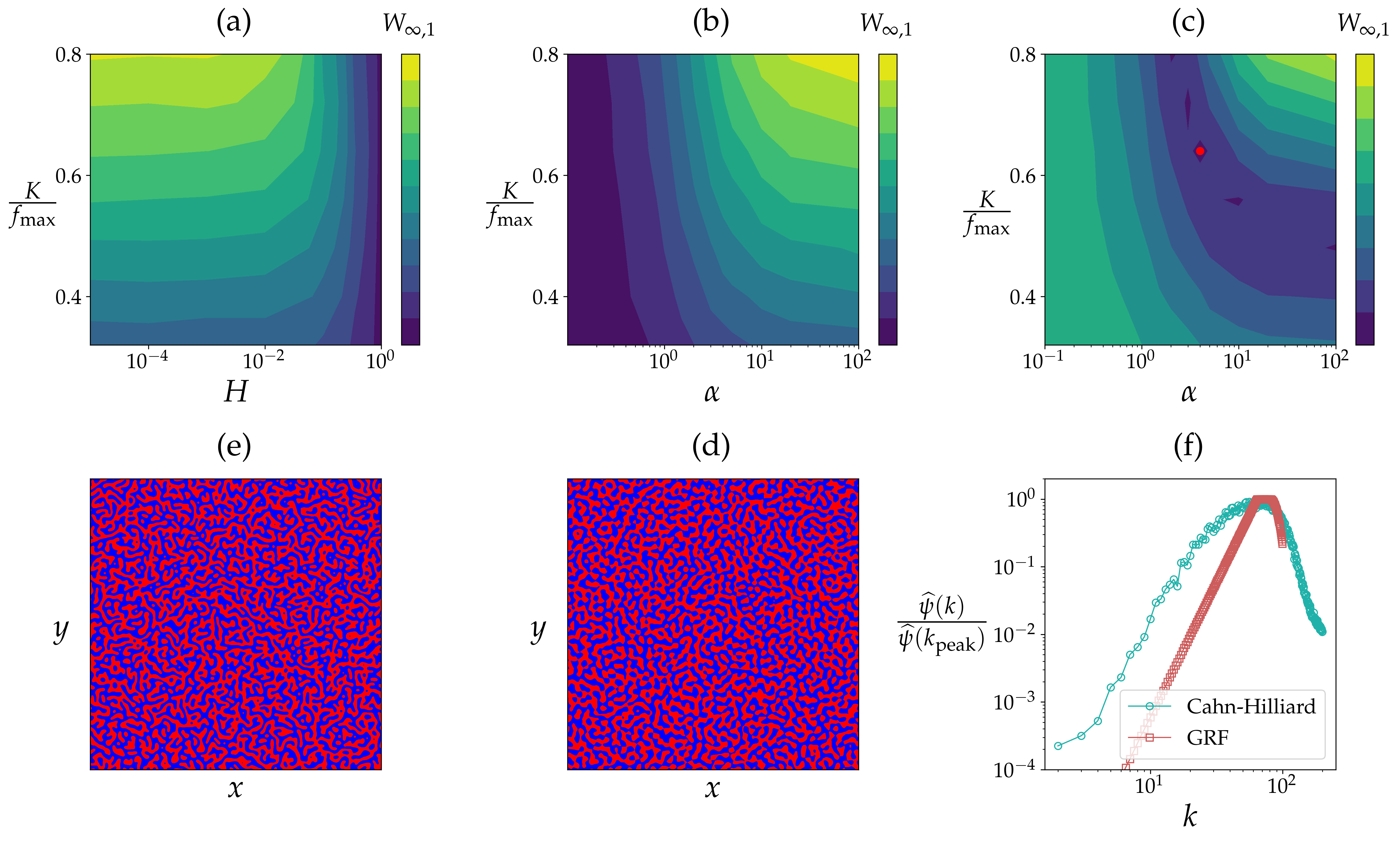}
    \caption{Quantifying differences between patterns via persistent homology via the distance $W_{\infty,1}$ between persistent diagrams of reference arrangements I and GRFs II. (a) I: random field; II: GRF with varying $H$ and $K$ with $\alpha=100$.(b)  I: random field; II: GRF with varying $\alpha$ and $K$ with $H=0$. (c) I: solution of the Cahn--Hilliard equation; II: GRF with varying $\alpha$ and $K$ with $H=0$. The red point indicates the pair $(\alpha, K)$ for which the minimum has been reached. (d) and (e) show the corresponding fields to the red point, namely the solution of the Cahn-Hilliard equation and the GRF with the smallest topological distance from it, respectively, the latter interpreted as a two-phase system via $\chi(\varphi(\mathbf{x})>0)$. Their spectral densities are shown in panel (f).}
    \label{fig:wasserstein_distances}
\end{figure}

We computed the Wasserstein distance $W_{\infty,1}$ between the persistence diagram of a random field and those of various fields featuring different values $\alpha$, $H$, and $K$ (generated as above). The results are summarized in Figure~\ref{fig:wasserstein_distances}(a) and \ref{fig:wasserstein_distances}(b). In the former, $\alpha = 100$ while varying $K$ and $H$ and in the latter, $H = 0$ while varying $\alpha$ and $K$ (approximating ideal HU character). These results demonstrate that information retained in persistence diagrams correlates with changes in the (global) spectral density. This result extends the statement proposed for HU point clouds \cite{salvalaglio2024persistent} to scalar fields: the HU character can be statistically determined by considering the distributions of its local geometrical/topological features. We note that a considerable change of the Wasserstein distance when varying $H \in [10^{-2}, 1]$ is obtained. This indicates a change in the local arrangements when approaching the nearly-HU state. For nearly-HU patterns ($H<10^{-2}$), minor changes can still be observed, especially for systems with large $K$ values, while no further changes are observed for $H<10^{-3}$. For the size considered here, this further supports considering patterns exhibiting $H<10^{-2}$ as belonging to the class of HU systems, although not formally satisfying eq.~\eqref{eq:def_field_hypruniformity}.

By exploiting the information in the persistence diagram, specific features of the spectral density of the corresponding field can also be identified. In other words, global properties can be reconstructed leveraging the distribution of local features as encoded in the persistence diagrams. In Figure~\ref{fig:wasserstein_distances}(c), we show the Wasserstein distance between the persistence diagrams of a solution of the Cahn-Hilliard equation, obtained as in Sect.~\ref{sec:CH}, and different fields for various $\alpha$, $K$, and $H=0$. The minimal value is obtained for $\alpha \sim 4$ and $K\sim0.64$ (red point in Figure~\ref{fig:wasserstein_distances}(c)). This GRF is the most similar to the solution of the Cahn-Hilliard equation in terms of the statistics over their local arrangements. Such a solution of the Cahn--Hilliard equation and the most similar GRF are illustrated in Figures~\ref{fig:wasserstein_distances}(d), (e), respectively. These patterns exhibit some morphological differences. Indeed, patterns obtained by the Cahn--Hilliard equation do not exactly belong to those described by the GRF parameterization via eqs.~\eqref{eq:GRF} and \eqref{eq:GRFparameters}. Nonetheless, they show parameters $\alpha$, $H$, and peaks of the spectral density, related to $K$ in eq. \eqref{eq:GRFparameters}, in the range of the considered GRFs. Figure \ref{fig:wasserstein_distances}(f) shows that the scaling $\widehat{\psi}(k) \sim k^4$ is analogous in both patterns while $K$ in the selected GRF well corresponds to the frequency $k_{\rm peak}$ of the considered solution of the Cahn--Hilliard equation. We thus explicitly obtained that global features of different patterns can be analyzed, compared, and, in principle, reconstructed by leveraging the distribution of arrangements as characterized by the considered persistence diagrams.

\subsection{Inferring hyperuniformity in GRFs}
\label{sec:machinel_learning}

Figure~\ref{fig:wasserstein_distances} already demonstrates that persistence diagrams of HU fields can be exploited to determine quantitatively their global characteristic through comparison with a reference. As shown for point clouds, such an inverse determination of HU characteristics can be generalized with the aid of machine learning \cite{Milor2025inferring}. 

The possibility of devising inverse HU-PH relations implies that the persistence diagrams stemming from different HU configurations are different from each other and, on a more general level, are different from those of non-HU origins. As a proof of concept, we consider the latter case, namely the distinction between HU and non-HU patterns in the GRFs defined in Eqs.~\eqref{eq:GRF} and \eqref{eq:GRFparameters}. Following \cite{Milor2025inferring}, we trained a neural network with three hidden layers, the first two with $128$ neurons, the last one with $256$ neurons. We process the persistence diagram $\mathcal{P}_k=\{(b_i, d_i)\}$ with the binning method, i.e. by dividing a certain subset of $\mathbb{R}^2$ space containing the diagram $\{(b_i, d_i-b_i)\}$ in $23\times23$ identical bins, the number of points inside of each bin resulting in a feature vector. For $\mathcal{P}_0$ and $\mathcal{P}_1$, this subset was respectively set to be $[-15, 8]\times[0, 15]$ and $[-8, 15]\times[0, 15]$. Finally, the accuracy results were obtained by averaging $30$ randomly initialized $5$-fold cross-validations. 

Following the results from Sec. \ref{sec:distance}, we define as HU those fields for which $H\leq 10^{-2}$. Since $\widetilde{H}$ tends to overestimate the value of $H$, as mentioned in Sect.~\ref{sec:HU}, we consider the pattern to be HU if $\widetilde{H}<0.011$. The accuracy of detection of the HU character is presented in Table~\ref{tab:accuracy_hyperuniformity}, which exceeds $97\%$ for all the analyzed patterns (HU and non-HU). The persistence diagram, computed using the isoline filtration and its binning, thus constitutes a comprehensive and exploitable descriptor for hyperuniformity. In other words, relations exist between HU characters and statistics over local geometrical arrangements, which can be exploited in numerical and data-driven approaches. Importantly, point clouds in persistent diagrams are highly dimensional descriptors, even more than those classically combined with ML methods, but can be tailored and coarsened via different binning, as already considered here. The proof of concept presented here may be used to efficiently screen large datasets of GRFs. It is worth mentioning that, relying solely on a training set including GRFs, this inverse approach cannot be exploited as a general tool for classifying hyperuniformity in any scalar field. However, we expect that using an extended parametrization of the fields in the training set, this approach can be extended and generalized for this purpose.

\begin{table}
\caption{\label{tab:accuracy_hyperuniformity} Accuracy in classifying HU and non-HU patterns using a neural network surrogate model trained on relationships between GRF and persistence diagrams.}
    \centering
    \footnotesize
    \setlength{\tabcolsep}{6pt}
    \renewcommand{\arraystretch}{2.5}
    \begin{tabular}{@{}l||lll}
        \hline
        \bf set & all patterns & HU patterns & Non-HU patterns\\
        \hline
        \hline
        \bf accuracy & \makecell{97.3\%\\($\pm$1.0\%)} & \makecell{97.2\%\\($\pm$1.6\%)} & \makecell{97.3\%\\($\pm$1.3\%)}\\
    \end{tabular}
\end{table}

\section{Conclusions}
\label{sec:conclusions}

We proposed a topological approach to numerically characterize disordered, correlated scalar fields, focusing on those that form in two-phase media and exhibit hyperuniformity. This approach exploits a topological data analysis technique, namely persistent homology, taking as input a filtration based on the signed distance from a representative field isoline. Differences between arrangements realized in the patterns are then quantified via Wasserstein distances between persistence diagrams.  

We applied this approach to study the topological features of patterns generated by the Cahn–Hilliard equation. We showed that these features converge numerically with $\epsilon \to 0$. In addition, we introduced a novel numerical analysis of how finite $\epsilon$ affects the self-similarity of evolving patterns in this model. Only for $\epsilon \to 0$, the patterns become self-similar, and the topological analysis confirms convergence even across different morphologies. For larger $\epsilon$, distances between persistence diagrams quantify deviations from self-similarity.

By generalizing the description of HU fields through GRFs, we showed that the global HU character systematically correlates with the distribution of local arrangements and topological features. Topological distances based on persistent homology enable the quantification of similarities and differences across a wide range of HU and non-HU fields. These distances can also be used to identify the closest match to a given pattern among generated ones and to reconstruct its global features. This has been illustrated for the HU characteristics of the solution of the Cahn-Hilliard equation. Finally, we reported on how information contained in a persistence diagram can serve as a descriptor for a simple surrogate classification model of GRFs.

The proposed topological approach has been applied to solutions of the Cahn-Hilliard equation and HU GRFs, but it is not limited to these specific applications. Similar analyses can be extended and applied to other two-phase systems, interfaces, and free boundary problems. Of particular interest from an applicational point of view are extensions to anisotropic and inhomogeneous systems \cite{Kumar2020}, as well as two-phase media with spatially varying properties. Additionally, promising applications include the use of data obtained via persistent homology as descriptors for designing disordered, correlated patterns.

\vspace{30pt}
\noindent
\textbf{Data Availability.} Data and software used in this study will be made available in open repositories (GitLab, Zenodo) upon acceptance for publication.


\begin{ack}
Computing resources have been provided by the Center for Information Services and High-Performance Computing (ZIH), the NHR Center of TU Dresden. This center is jointly supported by the Federal Ministry of Education and Research and the state governments participating in the NHR (\href{www.nhr-verein.de/unsere-partner}{www.nhr-verein.de/unsere-partner}).
\end{ack}

\begin{funding}
This work was supported by the German Research Foundation (DFG) -- Project numbers: 417223351 (Research Unit FOR3013), 493401063 (Research Training Group (GRK) 2868 D$^3$), 556185784, and by EPSRC -- Projects: EP/Y028872/1, International Oxford/Max Planck Institute collaboration grant EP/Z531224/1.
\end{funding}


\providecommand{\noopsort}[1]{}\providecommand{\singleletter}[1]{#1}%

\end{document}